\documentclass[aps,prl,preprint,superscriptaddress,showpacs]{revtex4-1}
\usepackage{graphicx}
\usepackage{amssymb, amsmath}
\usepackage[usenames]{color}
\usepackage{url}
\begin{document}

\title{Emergence of a Chern-insulating state from a semi-Dirac dispersion}

\author{Huaqing Huang}
\affiliation{Department of Physics and State Key Laboratory of
Low-Dimensional Quantum Physics, Tsinghua University, Beijing 100084,
China}
\affiliation{Department of Physics and Astronomy, Rutgers University,
Piscataway, New Jersey 08854-0849, USA}

\author{Zhirong Liu}
\affiliation{College of Chemistry and Molecular Engineering, Peking
University, Beijing 100871, China}

\author{Hongbin Zhang}
\affiliation{Department of Physics and Astronomy, Rutgers University,
Piscataway, New Jersey 08854-0849, USA}

\author{Wenhui Duan}
\affiliation{Department of Physics and State Key Laboratory of
Low-Dimensional Quantum Physics, Tsinghua University, Beijing 100084,
China}
\affiliation{Collaborative Innovation Center of Quantum Matter, Tsinghua
University, Beijing 100084, China}
\affiliation{Institute for Advanced Study, Tsinghua University, Beijing
100084, China}

\author{David Vanderbilt}
\affiliation{Department of Physics and Astronomy, Rutgers University,
Piscataway, New Jersey 08854-0849, USA}

\date{\today}

\begin{abstract}
A Chern insulator (quantum anomalous Hall insulator) phase is
demonstrated to exist in a typical semi-Dirac system, the TiO$_2$/VO$_2$
heterostructure.  By combining first-principles calculations with
Wannier-based tight-binding model, we calculate the Berry curvature
distribution, finding a Chern number of $-$2 for the valence bands,
and demonstrate the existence of gapless chiral edge states, ensuring
quantization of the Hall conductivity to $2e^2/h$. A new semi-Dirac model,
where each semi-Dirac cone is formed by merging three conventional Dirac
points, is proposed to reveal how the nontrivial topology with finite
Chern number is compatible with a semi-Dirac electronic spectrum.
\end{abstract}

\pacs{73.20.-r,73.21.-b,73.43.-f}
%73.20.At,
\maketitle

A Chern insulator is a two-dimensional (2D) magnetic insulator
with a quantized anomalous Hall conductivity $Ce^2/h$,
where $C$ is an integer topological index known as the Chern number
\cite{Haldane}. These systems, also known as quantum anomalous Hall
(QAH) insulators, have attracted a great deal of interest, in part
because of their gapless chiral edge states which enable
dissipationless transport, with potential applications in electronic
devices \cite{edge}. So far, several systems have been proposed,
notably magnetically doped topological insulators \cite{CXLiu,YuScience}
leading to recent experimental confirmation \cite{xue}, but also
honeycomb \cite{QiaoZhenhua,XiaoDi,Hongbin,LiuFeng,*LiuFeng2,Kevin,wucongjun}
or square lattices \cite{EuOCdO,EuOGdN,double-perovskite} formed by
transition-metal and heavy-metal ions. The essential ingredients
are the spontaneous breaking of time-reversal (TR) symmetry,
as by the formation of a ferromagnetic state, and the presence
of spin-orbit coupling (SOC), which generates the net Berry
curvature needed for a nonzero Chern number.  Here we focus on one
interesting class of proposals involving 2D systems that would be
half-semimetals in the absence of SOC, with the Fermi energy pinned
at one or more Dirac points in one spin channel while the other
channel is gapped.  If the application of SOC gaps the Dirac points
to produce conical avoided crossings instead, a nonzero Chern
number can result, as proposed for example for a triphenyl-manganese
(Mn$_2$C$_{18}$H$_{12}$) system \cite{LiuFeng,*LiuFeng2}.

Recently, calculations on TiO$_2$/VO$_2$ multilayer structures
led to a proposal \cite{victorPRL,victorPRB} for a new kind of
half-semimetal in which, in the absence of SOC, the bands in the
ungapped spin channel have a \textit{semi-Dirac} dispersion, i.e.,
a quadratic rather than a linear dispersion in one direction.
This raises the interesting question whether such a system can
also provides a route to a Chern-insulating state
when SOC is included.

In this work we answer this question.  First, we distinguish
between two qualitatively different types of semi-Dirac
structures. The one we denote as ``type-I'' was proposed in
Ref.~\cite{Banerjee}, but we find that it cannot lead to a
QAH state.  Instead, we find that the TiO$_2$/VO$_2$ multilayer
structure~\cite{victorPRL,victorPRB} is described by a new
``type-II'' semi-Dirac cone structure, which does lead to
a Chern-insulating state when SOC is turned on.  Moreover,
we clarify that the type-II structure is not protected by
symmetry, and (in the absence of SOC) will generically transform
into one or three Dirac nodes
in the absence of fine tuning. With SOC, we predict that the
TiO$_2$/VO$_2$ heterostructure is a QAH insulator with a Chern
number of $-2$, demonstrating a new route to the formation of
a Chern-insulating state in 2D.

\paragraph{Dirac and semi-Dirac cones.}
A general effective two-band Hamiltonian in 2D can be given as
\begin{equation}
H(\mathbf{k}) = \mathbf{h}(\mathbf{k})\cdot \vec{\sigma},
\label{Ham2D}
\end{equation}
where $\vec{\sigma}=(\sigma_x,\sigma_y,\sigma_z)$ are Pauli matrices.
In 2D $\bf k$-space a form like
\begin{equation}
\mathbf{h}(\mathbf{k})=(v_{\rm F}k_x, v_{\rm F}k_y, 0)
\label{Dirac}
\end{equation}
describes a massless Dirac cone structure, appropriate
to the case that SOC is absent,
where $v_{\rm F}$ is the Fermi velocity.
SOC is taken into account by adding a mass term
such that
$\mathbf{h}(\mathbf{k})=(v_{\rm F}k_x, v_{\rm F}k_y, m_z)$,
opening an energy gap of $2|m_z|$ at the Dirac point.
The Chern number is determined by
integrating the Berry curvature $\Omega(\mathbf{k})$ over the
2D Brillouin zone (BZ), but for weak SOC the dominant contributions
will come from Berry fluxes $\Phi=\pm\pi$ concentrated near the avoided
crossings.  This follows from the well-known form of the Berry
curvature for the two-band model of Eq.~(\ref{Ham2D}), which yields
\begin{equation}
\Phi=\int d\mathbf{k} \; \Omega(\mathbf{k})=\int
d\mathbf{k} \; \frac{\mathbf{h}}{2|\mathbf{h}|^3}\cdot(\frac{\partial
\mathbf{h}}{\partial k_x}\times\frac{\partial \mathbf{h}}{\partial  k_y}).
\label{Phi-def}
\end{equation}
Hence each gapped (massive) Dirac cone makes a contribution of
$\pm\frac{1}{2}$ to the total Chern number $C$, which is therefore
determined by summing over all the massive Dirac cones at
the Fermi level.

On the other hand, a semi-Dirac spectrum is a peculiar energy dispersion
in which quasiparticles behave as massless along one principal axis
but as massive fermions along the perpendicular direction.
Recently this novel spectrum was
observed in a multilayer (TiO$_2$)$_m$/(VO$_2$)$_n$ nanostructure ($m\geq5;
n=3$ or 4) by Pardo \textit{et al.}\ using first-principles calculations
\cite{victorPRL,victorPRB}.
They proposed that the semi-Dirac electronic
spectrum could be described by Eq.~(\ref{Ham2D}) with
\cite{Banerjee,PhysRevB.80.153412,*universal,PhysRevB.86.075124}
\begin{equation}
\mathbf{h}(\mathbf{k})=\left(\frac{k_x^2}{2m},v_{\rm F}k_y,0\right)
\label{TypeI}
\end{equation}
or a similar expression.  We refer to this model as a ``type-I'' semi-Dirac
model to distinguish it from the ``type-II'' model we propose below.
Obviously the dispersion is massless (linear) along the $k_y$ direction but
massive (quadratic) along $k_x$, satisfying the definition of
a semi-Dirac cone.  Since Eq.~(\ref{TypeI}) is an even function of $k_x$,
it is evident that the Berry curvature $\Omega(\mathbf{k})$ is an odd
function of $k_x$ according to Eq.~(\ref{Phi-def}).  Hence the Chern number,
given by the integral of $\Omega(\mathbf{k})$ over the vicinity of
the semi-Dirac point, is zero after adding a mass term $H^\prime=m_z \sigma_z$
to open an energy gap.  This is consistent with the observation that
the semi-Dirac point under Eq.~(\ref{TypeI}) carries a zero Berry flux
\cite{PhysRevB.80.153412,*universal}, in contrast with conventional Dirac
points which provide a Berry flux of $\pm\pi$.

We therefore conclude that type-I semi-Dirac cones have a trivial
topology with Chern number $C=0$ and thus do not lead to a Chern-insulating state.
Nevertheless, by using first-principles calculations and a Wannier-based
tight-binding analysis, we find that SOC does turn the
(TiO$_2$)$_m$/(VO$_2$)$_n$ into a Chern insulator.  We first present
the results and then explain them.

\paragraph{Methods.}
The first-principles electronic structure calculations are performed
within the framework of density functional theory (DFT) \cite{DFT}
as implemented in the QUANTUM-ESPRESSO package \cite{QE} with the
plane-wave pseudopotential method, and in the VASP package \cite{VASP} with
the projector augmented-wave method. We adopted the Perdew-Burke-Ernzerhof
generalized gradient approximation (GGA) exchange-correlation
functional \cite{PBE}.  The kinetic energy cutoff is fixed to be 500
eV and a $\Gamma$-centered $8\times8\times1$ $\mathbf{k}$-point mesh
is used in all cases.  We treat the open-shell $3d$ orbitals
by adding an effective Hubbard $U$ correction of 3.4 eV on V and Ti
atoms within the GGA+U approach \cite{HubbardU,HubbardU2}.
Structural relaxations are carried out without SOC, and then the
electronic structure is computed twice, once with and once without SOC.
Ferromagnetic ordering is found to be energetically favored in both
cases, and the ordering is along the \textit{z} axis when SOC is
present.  We use Wannier
interpolation based on maximally localized Wannier functions (MLWFs)
to calculate the Berry curvature and the anomalous Hall conductivity,
which requires a very dense $\mathbf{k}$-point grid in
the BZ \cite{wannier1,*wannier2,wannier90}.

\begin{figure}
\includegraphics[width =\columnwidth]{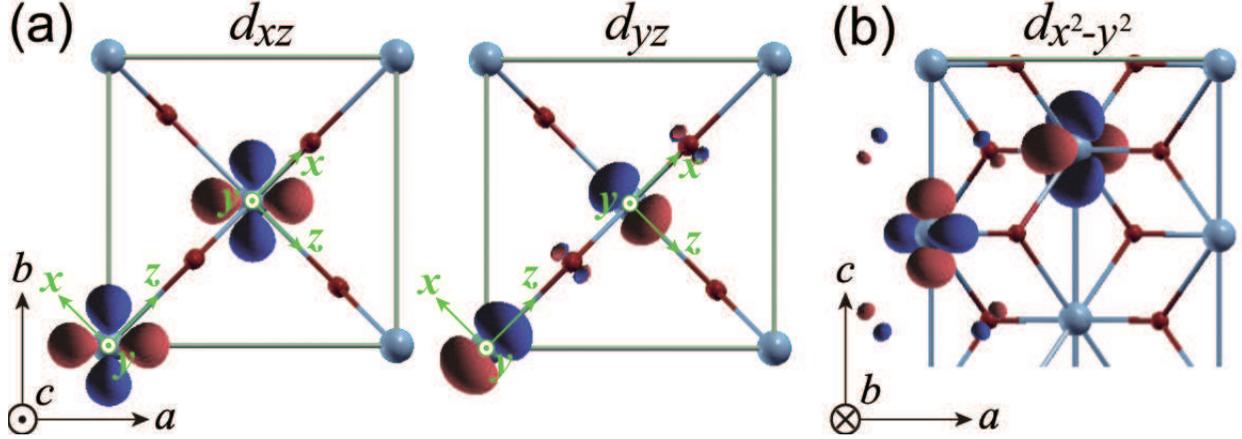}%
\caption{\label{iso}(Color online) Amplitude isosurfaces of
atom-centered $d$-like maximally localized Wannier functions (red
for positive values and blue for negative). (a) $d_\perp$ ($yz$
and $xz$) and (b) $d_\parallel$ ($x^2-y^2$) at different V sites in
the (TiO$_2$)$_5$/(VO$_2$)$_3$ superlattice. Local coordinate systems
are shown in green; local $z$ and $y$ axes are along the
in-plane O-V-O chain and the global $c$ axis respectively.
}
\end{figure}

\paragraph{Chern insulator behavior of the (TiO$_2$)$_5$/(VO$_2$)$_3$ system.}
We focus on the (TiO$_2$)$_m$/(VO$_2$)$_n$ system with
$n\!=\!3$ and $m\!=\!5$; because TiO$_2$ is strongly insulating,
the latter is enough to effectively separate the VO$_2$ trilayers
into isolated 2D systems, in which the energy bands close to the Fermi level
are dominated by the $3d$ states of V.
Since there are two kinds of VO$_6$ octahedra whose in-plane O-V-O
chains are perpendicular to each other, we align our local axes
differently on the two V sites as shown in Fig.~\ref{iso}(a).
Because of the distortion of the VO$_6$ octahedra away from cubic symmetry,
the triply degenerate $t_{2g}$ orbtials of V ions split into two doubly
degenerate $d_\perp$ orbtitals ($yz$ and $xz$) and one $d_\parallel$
orbital ($x^2-y^2$)~\cite{goodenough}. We projected the Bloch wavefunctions
onto these local orbitals to get MLWFs, which serve as an ideal basis
for further analyses. As shown in Fig.~\ref{iso}, The MLWFs keep the shape
and symmetry of local atomic $d$ orbitals.

\begin{figure*}
  \includegraphics[width =\textwidth]{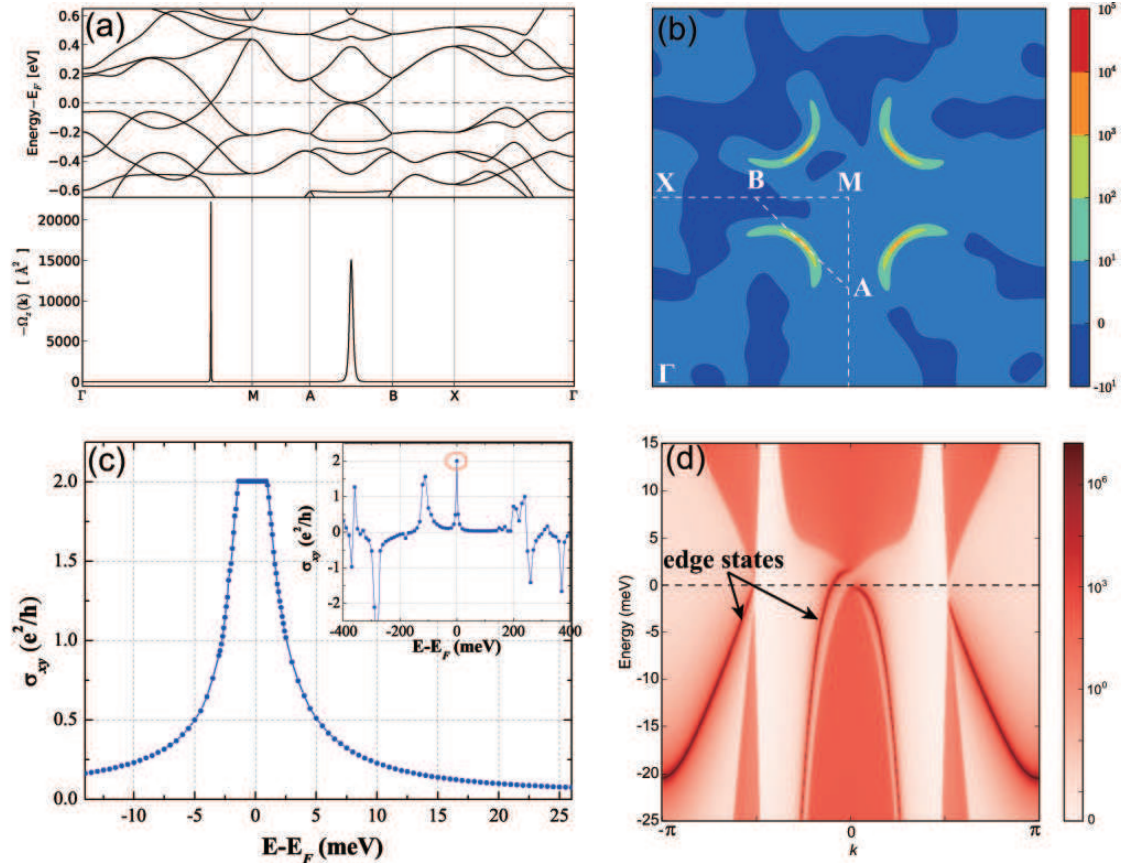}\\%2\columnwidth
  \centering
  \caption{\label{berry}(Color online) (a) Band structure and Berry
  curvature $-\Omega(\mathbf{k})$ along and perpendicular to the diagonal
  of the BZ. (b) Berry curvature $-\Omega(\mathbf{k})$ in the entire
  BZ. (c) Anomalous Hall conductivity $\sigma_{xy}$ plotted with respect
  to the position of the Fermi energy $\mathrm{E}_F$. The inset shows
  $\sigma_{xy}$ with $\mathrm{E}_F$ varying from $-400$ to $400$ meV.
  (d) Energy and momentum
  dependence of the LDOS at the edge along the [11] direction. Note that
  the color maps are logarithmic; deep blue patches in (b) are
  numerical artifacts.}
\end{figure*}

Based on these atom-centered \textit{d}-like MLWFs, we calculate the
Wannier-interpolated energy bands close to the Fermi level~\cite{Yates},
shown in the top panel of Fig.~\ref{berry}(a). The Wannier-interpolated energy
bands are in excellent agreement with the DFT results (not shown).
The semi-Dirac character is clearly apparent, with two bands crossing
linearly along the $\Gamma$--M line and quadratically along
A--B (perpendicular to
$\Gamma$--M). When SOC is considered in our calculation, %the two bands
an avoided crossing occurs and a small band gap of $\sim$2.5 meV opens at the
crossing point. To investigate the topological nature of this system,
we calculate the Berry curvature $\Omega(\mathbf{k})$ of all states
below the Fermi level using \cite{WangXinjie}
\begin{equation}
\Omega(\mathbf{k})=-\sum_{n<E_F}\sum_{m\neq n} {\rm 2Im}
\frac{\langle\psi_{n\textbf{k}}|v_x|\psi_{m\textbf{k}}\rangle
\langle\psi_{m\textbf{k}}|v_y|\psi_{n\textbf{k}}\rangle}
{(\varepsilon_{m\textbf{k}}-\varepsilon_{n\textbf{k}})^2},
\label{Omega}
\end{equation}
where $\psi_{n \rm \textbf{k}}$ is the spinor Bloch wavefunction of band
$n$ with corresponding eigenenergy $\varepsilon_{n \rm \textbf{k}}$,
and ${\bf v}=(v_x,v_y)$ is the velocity operator.
In the bottom panel of Fig.~\ref{berry}(a) the Berry curvature is plotted along the
same \textit{k}-path. The large peaks between the $\Gamma$ and M
points arise where the conduction and valence bands are nearly
degenerate and only weakly
split by SOC, giving rise to small denominators in Eq.~(\ref{Omega})
and hence a large contribution to $\Omega(\mathbf{k})$.  The peaks
show strong anisotropy, being much sharper along $\Gamma$--M than
along A--B.  The 2D plot in Fig.~\ref{berry}(b) makes it clear that these
avoided crossings give rise to four
banana-shaped peaks of Berry curvature located around the semi-Dirac
points.  Note that $\Omega(\mathbf{k})$
is not an odd function along the A--B direction, in contrast
with the prediction of Eq.~(\ref{TypeI}).  Since the four peaks are
related by fourfold rotations, there is no cancellation
among them; each contributes a Berry flux of $-\pi$, so that
the total Chern number is $-2$.
The system is therefore a Chern insulator.

We also plot the intrinsic anomalous Hall conductivity \cite{ahc} as a
function of Fermi energy in Fig.~\ref{berry}(c). As expected from the nonzero
Chern number, the anomalous Hall conductivity shows a quantized
Hall plateau at $\sigma_{xy}=-Ce^2/h$ when the Fermi level lies
inside the bulk band gap.  The width of the Hall plateau is about 2.5
meV, corresponding to the nontrivial bulk band gap. The anomalous Hall
conductivity decreases rapidly to zero when the Fermi level is outside
the band gap, as expected since only states near the nontrivial
bulk band gap, which is opened by the weak SOC, contribute strongly
to the Berry curvature [see Fig. 2(a,b)].

The existence of topologically protected chiral edge states is one of
the most important consequences of the QAH state. To further reveal the
nontrivial topological nature of the system, we calculate the edge
states of a semi-infinite (TiO$_2$)$_5$/(VO$_2$)$_3$ system with
its edge along the [11]
direction. Because the existence of chiral edge states is completely
determined by the bulk topology, here we build the tight-binding model
with SOC based on first-principles MLWFs from the bulk,
ignoring the effects of edge reconstruction. We apply an iterative method
\cite{lopez,*lopez2} to obtain the edge Green's function and the local
density of states, which is directly related to the imaginary part of
the Green's function. In Fig.~\ref{berry}(d) we can clearly see that two edges
states connect the valence and conduction bands. The appearance of
two chiral edge states is consistent with our calculated Chern
number $C=-2$, confirming the nontrivial topological nature of
this semi-Dirac system.

\paragraph{Type-II semi-Dirac model.}
The above results demonstrate that the (TiO$_2$)$_5$/(VO$_2$)$_3$
superlattice is undoubtedly a Chern insulator,
contrary to previous understandings that a semi-Dirac cone
structure should not result in a Chern-insulating state.
To resolve such a conflict, we propose a
``type-II'' semi-Dirac model with an effective $2 \times 2$ Hamiltonian
in the absence of SOC having the form of Eq.~(\ref{Ham2D}) with
\begin{equation}
\mathbf{h}(\mathbf{k})=
   \left(\frac{k_x^2}{2m}-v_{\rm F}k_y,\alpha k_xk_y,0\right)
\label{TypeII}
\end{equation}
or other equivalent expressions.  The resulting dispersion relation
satisfies the semi-Dirac character: it is linear along one direction
($k_x\!=\!0$) and quadratic perpendicular to this direction ($k_y\!=\!0$).
Hence the origin point $(0, 0)$ is a semi-Dirac point.  We then add
a SOC-induced mass term $H^\prime=m_z \sigma_z$ to open a gap at
the semi-Dirac point.

\begin{figure}
\includegraphics[width =\columnwidth]{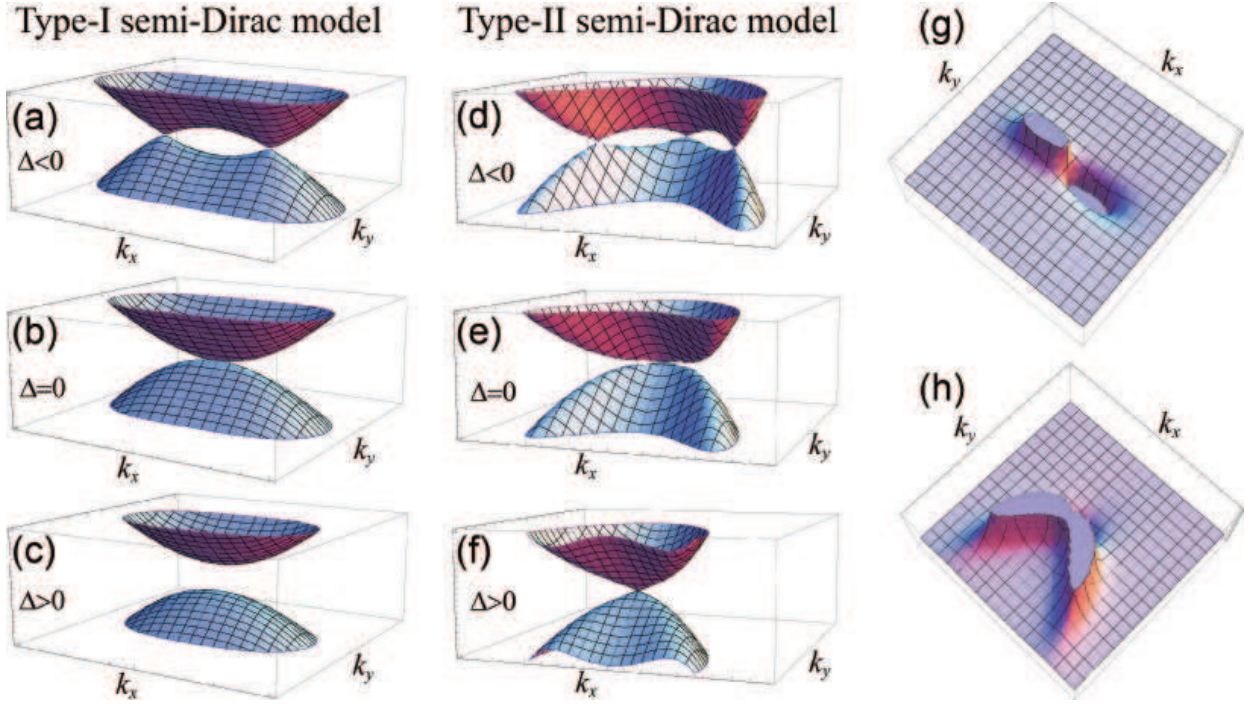}%
\caption{\label{bsr}(Color online) Evolution of band structures of
(a)-(c) type-I and (d)-(f) type-II semi-Dirac dispersions when
$h_x$ aquires a small perturbation $\Delta$.  Berry curvature
$-\Omega$ of (g) type-I and (h) type-II semi-Dirac models with
$\Delta=0$ when a SOC-induced mass term $H^\prime=m_z\sigma_z$
is added to open a small gap at the semi-Dirac point.}
\end{figure}

To understand the nature of the Berry curvature expected from
the type-I and type-II models of Eqs.~(\ref{TypeI}) and (\ref{TypeII}),
we plot band dispersions and Berry curvatures derived from
these models in Fig.~\ref{bsr}.
In particular, we investigate the evolution of the
crossing points
when $h_x$ aquires a small perturbation $\Delta$ in both models
(assuming $m>0$ without loss of generality).
In the type-I model this leads to the the fusion of two Dirac
points as $\Delta$ passes through zero.  As shown in Figs.~3(a-c),
when $\Delta<0$ there are two Dirac cones located at ${\bf k}=
(\pm\sqrt{-2m\Delta},0)$; these merge to form a semi-Dirac cone
at $\Delta\!=\!0$, and then a gap of magnitude $2\Delta$ appears
for $\Delta>0$.  Because the two Dirac points provide Berry
fluxes of opposite sign, the total Berry flux vanishes, as shown
for $\Delta\!=\!0$ in Fig.~\ref{bsr}(g).  Thus, the type-I
model is topologically trivial, and cannot be used to describe the
semi-Dirac structure observed for the (TiO$_2$)$_5$/(VO$_2$)$_3$
superlattice system.

For the type-II model, on the other hand, when $\Delta<0$ the
energy gap vanishes at \textit{three} Dirac points located at
${\bf k}=(0,\Delta/v_F)$ and $(\pm \sqrt{-2m\Delta},0)$,
as shown in Fig.~\ref{bsr}(d). As $\Delta$ increases, these approach each
other and merge at $\Delta\!=\!0$ into a single semi-Dirac point
at ${\bf k}=(0,0)$ [Fig.~\ref{bsr}(e)]. For $\Delta>0$, only the touching
at ${\bf k}=(0,\Delta/v_F)$ remains; the other two points
disappear (their coordinates formally become imaginary)
[Fig.~\ref{bsr}(f)]. Therefore, the type-II semi-Dirac model
of Eq.~(\ref{TypeII}) can be viewed as the consequence of the merging of
three Dirac points to produce a single Dirac point.
The distribution of the Berry flux produced by the type-II model
at $\Delta\!=\!0$ is shown in Fig.~\ref{bsr}(h).  Significantly, the
Berry curvature near the semi-Dirac point has the banana shape
shown in Fig.~\ref{bsr}(h), very similar to the DFT results in Fig.~\ref{berry}(b).

For one sign of $m_z$, the total Berry flux contributed by the
type-II model is $+\pi$ independent of $\Delta$; the two Dirac
points at $(\pm \sqrt{-2m\Delta},0)$ contribute $+\pi$ when
$\Delta<0$, while the one at ${\bf k}=(0,\Delta/v_F)$ switches
from $-\pi$ to $+\pi$ as $\Delta$ crosses through zero.
For the other sign of $m_z$, all these are reversed.  Since there are four
symmetry-related semi-Dirac points in the entire BZ, the total Chern number
is $\pm2$. The results of the type-II semi-Dirac model thus
agree very well with the DFT calculations of
(TiO$_2$)$_5$/(VO$_2$)$_3$ above, strongly suggesting that the
type-II model correctly describes this system.

To further verify this
conclusion, we conduct an effective $\mathbf{k}\cdot\mathbf{p}$
analysis on the first-principles data without SOC around the band crossing
point at $(k_c,k_c)$ and introduce a coordinate transformation
$\mathbf{q}=(q_1,q_2)$ with $q_1=(k_x+k_y)/2-k_c$ and $q_2=(k_x-k_y)/2$.
By employing the downfolding method \cite{downfold1,downfold2}
we obtain the $2 \times 2$ effective Hamiltonian \footnote{\label{fn}See
  Supplementary Material http://link.aps.org/supplemental/xxx for
  more computation details, the derivation of the $\mathbf{k}\cdot\mathbf{p}$ model, and
  a detailed analysis of the electronic structure.}
\begin{equation}
H_{\mathbf{k}\cdot\mathbf{p}} = \epsilon (\mathbf{q})\mathbb{I}_{2 \times 2} +
\mathbf{h(\mathbf{q})}\cdot \vec{\sigma},
\label{Hkp}
\end{equation}
where $\mathbb{I}_{2 \times 2}$ is the unit matrix.
Keeping only terms up to quadratic order in $\mathbf{q}$
we find $\mathbf{h(\mathbf{q})}=(Aq_1+Bq_2^2,Cq_2+Dq_1q_2,0)$.
The coefficient $C$ is small but non-zero, implying that the
dispersion along $q_2$ is almost quadratic, but contains a small
linear component.  We have checked that the presence of a nonzero
coefficient $C$ is not prevented by any symmetry.
Setting ${\bf h}\!=\!0$ we find three zero-gap Dirac points: a
real one at $q_1\!=\!q_2\!=\!0$ and two virtual ones at
$(0.20, \pm0.56i)$ (in units of $a^{-1}$, where $a$ is the
in-plane lattice constant).
These results map onto the $\Delta$-modified type-II model as
discussed following Eq.~(\ref{TypeII}), with $q_1=k_y-\Delta/v_{\rm F}$
and $q_2=k_x$, so we can identify $A\!=\!-v_{\rm F}$,
$B\!=\!1/2m$, $C\!=\!\alpha\Delta/v_{\rm F}$, and $D\!=\!\alpha$.
Putting in numbers, we find a small positive $\Delta$, placing
the system into the regime of Fig.~\ref{bsr}(f), but close to Fig.~\ref{bsr}(e).
If we artifically decrease $C$ and let it pass through zero,
the real and virtual Dirac points merge at $\mathbf{q}=(0,0)$ and
reemerge as three real ones, exactly as predicted in Fig.~\ref{bsr}(d) for
the type-II semi-Dirac model.

We therefore conclude that the (TiO$_2$)$_5$/(VO$_2$)$_3$
nanostructure is an anisotropic linear Dirac system that is
so close to a type-II semi-Dirac behavior that it can hardly
be distinguished from it. When the system is gapped by SOC,
the calculated Berry flux is $\Phi\approx-\pi$ and the Berry
curvatures are banana-shaped, independent of whether the linear
term of $q_2$ is exactly zero or not.\footnotemark[\value{footnote}] Hence the Chern
insulator phase is quite robust, even if the spectrum diverge
from the exact semi-Dirac dispersion due to external strain or other
perturbations.

\paragraph{Conclusion.}
We have found that the (TiO$_2$)$_5$/(VO$_2$)$_3$ nanostructure,
which exhibits a peculiar semi-Dirac-like
electronic structure in the absence of SOC, is a QAH
(Chern) insulator with $C\!=\!-2$ when SOC is included.
The calculated Berry curvature, anomalous
Hall conductivity and gapless edge states unambiguously identify
the nontrivial topological nature of this system. We propose a new
semi-Dirac model, which describes the semi-Dirac point as the merging
of three conventional Dirac points, to explain how a Chern-insulating state
is compatible with a semi-Dirac dispersion.  Our results
clarify when and how the QAH state can emerge from a
semi-Dirac structure, providing a new potential route
to the realization of such states in practical materials systems.

\begin{acknowledgments}
We thank Ivo Souza, Jianpeng Liu and Wendong Cao for valuable
discussions. D. V. is supported by NSF Grant DMR-10-05838.
H. H., Z. L. and W. D. are
supported by the National Natural Science Foundation of China (Grant
Nos. 21373015 and 11334006) and the Ministry of Science
and Technology of China (Grant Nos. 2011CB606405 and 2011CB921901).
\end{acknowledgments}

\paragraph{Note added}--
Recently, a QAH phase was also found in TiO$_2$/CrO$_2$ heterostructures
by Cai \textit{et al.} \cite{Cai}.  However, it exhibits nearly isotropic
Dirac cones, which is quite different from the semi-Dirac cones of
TiO$_2$/VO$_2$ in our work.

%\bibliography{SemiDirac}

\begin{thebibliography}{40}%
\makeatletter
\providecommand \@ifxundefined [1]{%
 \@ifx{#1\undefined}
}%
\providecommand \@ifnum [1]{%
 \ifnum #1\expandafter \@firstoftwo
 \else \expandafter \@secondoftwo
 \fi
}%
\providecommand \@ifx [1]{%
 \ifx #1\expandafter \@firstoftwo
 \else \expandafter \@secondoftwo
 \fi
}%
\providecommand \natexlab [1]{#1}%
\providecommand \enquote  [1]{``#1''}%
\providecommand \bibnamefont  [1]{#1}%
\providecommand \bibfnamefont [1]{#1}%
\providecommand \citenamefont [1]{#1}%
\providecommand \href@noop [0]{\@secondoftwo}%
\providecommand \href [0]{\begingroup \@sanitize@url \@href}%
\providecommand \@href[1]{\@@startlink{#1}\@@href}%
\providecommand \@@href[1]{\endgroup#1\@@endlink}%
\providecommand \@sanitize@url [0]{\catcode `\\12\catcode `\$12\catcode
  `\&12\catcode `\#12\catcode `\^12\catcode `\_12\catcode `\%12\relax}%
\providecommand \@@startlink[1]{}%
\providecommand \@@endlink[0]{}%
\providecommand \url  [0]{\begingroup\@sanitize@url \@url }%
\providecommand \@url [1]{\endgroup\@href {#1}{\urlprefix }}%
\providecommand \urlprefix  [0]{URL }%
\providecommand \Eprint [0]{\href }%
\providecommand \doibase [0]{http://dx.doi.org/}%
\providecommand \selectlanguage [0]{\@gobble}%
\providecommand \bibinfo  [0]{\@secondoftwo}%
\providecommand \bibfield  [0]{\@secondoftwo}%
\providecommand \translation [1]{[#1]}%
\providecommand \BibitemOpen [0]{}%
\providecommand \bibitemStop [0]{}%
\providecommand \bibitemNoStop [0]{.\EOS\space}%
\providecommand \EOS [0]{\spacefactor3000\relax}%
\providecommand \BibitemShut  [1]{\csname bibitem#1\endcsname}%
\let\auto@bib@innerbib\@empty
%</preamble>
\bibitem [{\citenamefont {Haldane}(1988)}]{Haldane}%
  \BibitemOpen
  \bibfield  {author} {\bibinfo {author} {\bibfnamefont {F.~D.~M.}\
  \bibnamefont {Haldane}},\ }\href {\doibase 10.1103/PhysRevLett.61.2015}
  {\bibfield  {journal} {\bibinfo  {journal} {Phys. Rev. Lett.}\ }\textbf
  {\bibinfo {volume} {61}},\ \bibinfo {pages} {2015} (\bibinfo {year}
  {1988})}\BibitemShut {NoStop}%
\bibitem [{\citenamefont {Halperin}(1982)}]{edge}%
  \BibitemOpen
  \bibfield  {author} {\bibinfo {author} {\bibfnamefont {B.~I.}\ \bibnamefont
  {Halperin}},\ }\href {\doibase 10.1103/PhysRevB.25.2185} {\bibfield
  {journal} {\bibinfo  {journal} {Phys. Rev. B}\ }\textbf {\bibinfo {volume}
  {25}},\ \bibinfo {pages} {2185} (\bibinfo {year} {1982})}\BibitemShut
  {NoStop}%
\bibitem [{\citenamefont {Liu}\ \emph {et~al.}(2008)\citenamefont {Liu},
  \citenamefont {Qi}, \citenamefont {Dai}, \citenamefont {Fang},\ and\
  \citenamefont {Zhang}}]{CXLiu}%
  \BibitemOpen
  \bibfield  {author} {\bibinfo {author} {\bibfnamefont {C.-X.}\ \bibnamefont
  {Liu}}, \bibinfo {author} {\bibfnamefont {X.-L.}\ \bibnamefont {Qi}},
  \bibinfo {author} {\bibfnamefont {X.}~\bibnamefont {Dai}}, \bibinfo {author}
  {\bibfnamefont {Z.}~\bibnamefont {Fang}}, \ and\ \bibinfo {author}
  {\bibfnamefont {S.-C.}\ \bibnamefont {Zhang}},\ }\href {\doibase
  10.1103/PhysRevLett.101.146802} {\bibfield  {journal} {\bibinfo  {journal}
  {Phys. Rev. Lett.}\ }\textbf {\bibinfo {volume} {101}},\ \bibinfo {pages}
  {146802} (\bibinfo {year} {2008})}\BibitemShut {NoStop}%
\bibitem [{\citenamefont {Yu}\ \emph {et~al.}(2010)\citenamefont {Yu},
  \citenamefont {Zhang}, \citenamefont {Zhang}, \citenamefont {Zhang},
  \citenamefont {Dai},\ and\ \citenamefont {Fang}}]{YuScience}%
  \BibitemOpen
  \bibfield  {author} {\bibinfo {author} {\bibfnamefont {R.}~\bibnamefont
  {Yu}}, \bibinfo {author} {\bibfnamefont {W.}~\bibnamefont {Zhang}}, \bibinfo
  {author} {\bibfnamefont {H.-J.}\ \bibnamefont {Zhang}}, \bibinfo {author}
  {\bibfnamefont {S.-C.}\ \bibnamefont {Zhang}}, \bibinfo {author}
  {\bibfnamefont {X.}~\bibnamefont {Dai}}, \ and\ \bibinfo {author}
  {\bibfnamefont {Z.}~\bibnamefont {Fang}},\ }\href {\doibase
  10.1126/science.1187485} {\bibfield  {journal} {\bibinfo  {journal}
  {Science}\ }\textbf {\bibinfo {volume} {329}},\ \bibinfo {pages} {61}
  (\bibinfo {year} {2010})}\BibitemShut {NoStop}%
\bibitem [{\citenamefont {Chang}\ and\ \citenamefont {\textit{et
  al.}}(2013)}]{xue}%
  \BibitemOpen
  \bibfield  {author} {\bibinfo {author} {\bibfnamefont {C.-Z.}\ \bibnamefont
  {Chang}}\ and\ \bibinfo {author} {\bibnamefont {\textit{et al.}}},\ }\href
  {\doibase 10.1126/science.1234414} {\bibfield  {journal} {\bibinfo  {journal}
  {Science}\ }\textbf {\bibinfo {volume} {340}},\ \bibinfo {pages} {167}
  (\bibinfo {year} {2013})}\BibitemShut {NoStop}%
\bibitem [{\citenamefont {Qiao}\ \emph {et~al.}(2010)\citenamefont {Qiao},
  \citenamefont {Yang}, \citenamefont {Feng}, \citenamefont {Tse},
  \citenamefont {Ding}, \citenamefont {Yao}, \citenamefont {Wang},\ and\
  \citenamefont {Niu}}]{QiaoZhenhua}%
  \BibitemOpen
  \bibfield  {author} {\bibinfo {author} {\bibfnamefont {Z.}~\bibnamefont
  {Qiao}}, \bibinfo {author} {\bibfnamefont {S.~A.}\ \bibnamefont {Yang}},
  \bibinfo {author} {\bibfnamefont {W.}~\bibnamefont {Feng}}, \bibinfo {author}
  {\bibfnamefont {W.-K.}\ \bibnamefont {Tse}}, \bibinfo {author} {\bibfnamefont
  {J.}~\bibnamefont {Ding}}, \bibinfo {author} {\bibfnamefont {Y.}~\bibnamefont
  {Yao}}, \bibinfo {author} {\bibfnamefont {J.}~\bibnamefont {Wang}}, \ and\
  \bibinfo {author} {\bibfnamefont {Q.}~\bibnamefont {Niu}},\ }\href {\doibase
  10.1103/PhysRevB.82.161414} {\bibfield  {journal} {\bibinfo  {journal} {Phys.
  Rev. B}\ }\textbf {\bibinfo {volume} {82}},\ \bibinfo {pages} {161414}
  (\bibinfo {year} {2010})}\BibitemShut {NoStop}%
\bibitem [{\citenamefont {Xiao}\ \emph {et~al.}(2011)\citenamefont {Xiao},
  \citenamefont {Zhu}, \citenamefont {Ran}, \citenamefont {Nagaosa},\ and\
  \citenamefont {Okamoto}}]{XiaoDi}%
  \BibitemOpen
  \bibfield  {author} {\bibinfo {author} {\bibfnamefont {D.}~\bibnamefont
  {Xiao}}, \bibinfo {author} {\bibfnamefont {W.}~\bibnamefont {Zhu}}, \bibinfo
  {author} {\bibfnamefont {Y.}~\bibnamefont {Ran}}, \bibinfo {author}
  {\bibfnamefont {N.}~\bibnamefont {Nagaosa}}, \ and\ \bibinfo {author}
  {\bibfnamefont {S.}~\bibnamefont {Okamoto}},\ }\href@noop {} {\bibfield
  {journal} {\bibinfo  {journal} {Nat. Commun.}\ }\textbf {\bibinfo {volume}
  {2}},\ \bibinfo {pages} {596} (\bibinfo {year} {2011})}\BibitemShut {NoStop}%
\bibitem [{\citenamefont {Zhang}\ \emph {et~al.}(2012)\citenamefont {Zhang},
  \citenamefont {Lazo}, \citenamefont {Bl\"ugel}, \citenamefont {Heinze},\ and\
  \citenamefont {Mokrousov}}]{Hongbin}%
  \BibitemOpen
  \bibfield  {author} {\bibinfo {author} {\bibfnamefont {H.}~\bibnamefont
  {Zhang}}, \bibinfo {author} {\bibfnamefont {C.}~\bibnamefont {Lazo}},
  \bibinfo {author} {\bibfnamefont {S.}~\bibnamefont {Bl\"ugel}}, \bibinfo
  {author} {\bibfnamefont {S.}~\bibnamefont {Heinze}}, \ and\ \bibinfo {author}
  {\bibfnamefont {Y.}~\bibnamefont {Mokrousov}},\ }\href {\doibase
  10.1103/PhysRevLett.108.056802} {\bibfield  {journal} {\bibinfo  {journal}
  {Phys. Rev. Lett.}\ }\textbf {\bibinfo {volume} {108}},\ \bibinfo {pages}
  {056802} (\bibinfo {year} {2012})}\BibitemShut {NoStop}%
\bibitem [{\citenamefont {Wang}\ \emph {et~al.}(2013)\citenamefont {Wang},
  \citenamefont {Liu},\ and\ \citenamefont {Liu}}]{LiuFeng}%
  \BibitemOpen
  \bibfield  {author} {\bibinfo {author} {\bibfnamefont {Z.~F.}\ \bibnamefont
  {Wang}}, \bibinfo {author} {\bibfnamefont {Z.}~\bibnamefont {Liu}}, \ and\
  \bibinfo {author} {\bibfnamefont {F.}~\bibnamefont {Liu}},\ }\href {\doibase
  10.1103/PhysRevLett.110.196801} {\bibfield  {journal} {\bibinfo  {journal}
  {Phys. Rev. Lett.}\ }\textbf {\bibinfo {volume} {110}},\ \bibinfo {pages}
  {196801} (\bibinfo {year} {2013})}\BibitemShut {NoStop}%
\bibitem [{\citenamefont {Hu}\ \emph {et~al.}(2014)\citenamefont {Hu},
  \citenamefont {Wang},\ and\ \citenamefont {Liu}}]{LiuFeng2}%
  \BibitemOpen
  \bibfield  {author} {\bibinfo {author} {\bibfnamefont {H.}~\bibnamefont
  {Hu}}, \bibinfo {author} {\bibfnamefont {Z.}~\bibnamefont {Wang}}, \ and\
  \bibinfo {author} {\bibfnamefont {F.}~\bibnamefont {Liu}},\ }\href@noop {}
  {\bibfield  {journal} {\bibinfo  {journal} {Nanoscale Res. Lett.}\ }\textbf
  {\bibinfo {volume} {9}},\ \bibinfo {pages} {690} (\bibinfo {year}
  {2014})}\BibitemShut {NoStop}%
\bibitem [{\citenamefont {Garrity}\ and\ \citenamefont
  {Vanderbilt}(2013)}]{Kevin}%
  \BibitemOpen
  \bibfield  {author} {\bibinfo {author} {\bibfnamefont {K.~F.}\ \bibnamefont
  {Garrity}}\ and\ \bibinfo {author} {\bibfnamefont {D.}~\bibnamefont
  {Vanderbilt}},\ }\href {\doibase 10.1103/PhysRevLett.110.116802} {\bibfield
  {journal} {\bibinfo  {journal} {Phys. Rev. Lett.}\ }\textbf {\bibinfo
  {volume} {110}},\ \bibinfo {pages} {116802} (\bibinfo {year}
  {2013})}\BibitemShut {NoStop}%
\bibitem [{\citenamefont {Zhang}\ \emph
  {et~al.}(2014{\natexlab{a}})\citenamefont {Zhang}, \citenamefont {Li},\ and\
  \citenamefont {Wu}}]{wucongjun}%
  \BibitemOpen
  \bibfield  {author} {\bibinfo {author} {\bibfnamefont {G.-F.}\ \bibnamefont
  {Zhang}}, \bibinfo {author} {\bibfnamefont {Y.}~\bibnamefont {Li}}, \ and\
  \bibinfo {author} {\bibfnamefont {C.}~\bibnamefont {Wu}},\ }\href {\doibase
  10.1103/PhysRevB.90.075114} {\bibfield  {journal} {\bibinfo  {journal} {Phys.
  Rev. B}\ }\textbf {\bibinfo {volume} {90}},\ \bibinfo {pages} {075114}
  (\bibinfo {year} {2014}{\natexlab{a}})}\BibitemShut {NoStop}%
\bibitem [{\citenamefont {Zhang}\ \emph
  {et~al.}(2014{\natexlab{b}})\citenamefont {Zhang}, \citenamefont {Wang},
  \citenamefont {Xu}, \citenamefont {Xu},\ and\ \citenamefont
  {Zhang}}]{EuOCdO}%
  \BibitemOpen
  \bibfield  {author} {\bibinfo {author} {\bibfnamefont {H.}~\bibnamefont
  {Zhang}}, \bibinfo {author} {\bibfnamefont {J.}~\bibnamefont {Wang}},
  \bibinfo {author} {\bibfnamefont {G.}~\bibnamefont {Xu}}, \bibinfo {author}
  {\bibfnamefont {Y.}~\bibnamefont {Xu}}, \ and\ \bibinfo {author}
  {\bibfnamefont {S.-C.}\ \bibnamefont {Zhang}},\ }\href {\doibase
  10.1103/PhysRevLett.112.096804} {\bibfield  {journal} {\bibinfo  {journal}
  {Phys. Rev. Lett.}\ }\textbf {\bibinfo {volume} {112}},\ \bibinfo {pages}
  {096804} (\bibinfo {year} {2014}{\natexlab{b}})}\BibitemShut {NoStop}%
\bibitem [{\citenamefont {Garrity}\ and\ \citenamefont
  {Vanderbilt}(2014)}]{EuOGdN}%
  \BibitemOpen
  \bibfield  {author} {\bibinfo {author} {\bibfnamefont {K.~F.}\ \bibnamefont
  {Garrity}}\ and\ \bibinfo {author} {\bibfnamefont {D.}~\bibnamefont
  {Vanderbilt}},\ }\href {\doibase 10.1103/PhysRevB.90.121103} {\bibfield
  {journal} {\bibinfo  {journal} {Phys. Rev. B}\ }\textbf {\bibinfo {volume}
  {90}},\ \bibinfo {pages} {121103} (\bibinfo {year} {2014})}\BibitemShut
  {NoStop}%
\bibitem [{\citenamefont {Zhang}\ \emph
  {et~al.}(2014{\natexlab{c}})\citenamefont {Zhang}, \citenamefont {Huang},
  \citenamefont {Haule},\ and\ \citenamefont {Vanderbilt}}]{double-perovskite}%
  \BibitemOpen
  \bibfield  {author} {\bibinfo {author} {\bibfnamefont {H.}~\bibnamefont
  {Zhang}}, \bibinfo {author} {\bibfnamefont {H.}~\bibnamefont {Huang}},
  \bibinfo {author} {\bibfnamefont {K.}~\bibnamefont {Haule}}, \ and\ \bibinfo
  {author} {\bibfnamefont {D.}~\bibnamefont {Vanderbilt}},\ }\href {\doibase
  10.1103/PhysRevB.90.165143} {\bibfield  {journal} {\bibinfo  {journal} {Phys.
  Rev. B}\ }\textbf {\bibinfo {volume} {90}},\ \bibinfo {pages} {165143}
  (\bibinfo {year} {2014}{\natexlab{c}})}\BibitemShut {NoStop}%
\bibitem [{\citenamefont {Pardo}\ and\ \citenamefont
  {Pickett}(2009)}]{victorPRL}%
  \BibitemOpen
  \bibfield  {author} {\bibinfo {author} {\bibfnamefont {V.}~\bibnamefont
  {Pardo}}\ and\ \bibinfo {author} {\bibfnamefont {W.~E.}\ \bibnamefont
  {Pickett}},\ }\href {\doibase 10.1103/PhysRevLett.102.166803} {\bibfield
  {journal} {\bibinfo  {journal} {Phys. Rev. Lett.}\ }\textbf {\bibinfo
  {volume} {102}},\ \bibinfo {pages} {166803} (\bibinfo {year}
  {2009})}\BibitemShut {NoStop}%
\bibitem [{\citenamefont {Pardo}\ and\ \citenamefont
  {Pickett}(2010)}]{victorPRB}%
  \BibitemOpen
  \bibfield  {author} {\bibinfo {author} {\bibfnamefont {V.}~\bibnamefont
  {Pardo}}\ and\ \bibinfo {author} {\bibfnamefont {W.~E.}\ \bibnamefont
  {Pickett}},\ }\href {\doibase 10.1103/PhysRevB.81.035111} {\bibfield
  {journal} {\bibinfo  {journal} {Phys. Rev. B}\ }\textbf {\bibinfo {volume}
  {81}},\ \bibinfo {pages} {035111} (\bibinfo {year} {2010})}\BibitemShut
  {NoStop}%
\bibitem [{\citenamefont {Banerjee}\ \emph {et~al.}(2009)\citenamefont
  {Banerjee}, \citenamefont {Singh}, \citenamefont {Pardo},\ and\ \citenamefont
  {Pickett}}]{Banerjee}%
  \BibitemOpen
  \bibfield  {author} {\bibinfo {author} {\bibfnamefont {S.}~\bibnamefont
  {Banerjee}}, \bibinfo {author} {\bibfnamefont {R.~R.~P.}\ \bibnamefont
  {Singh}}, \bibinfo {author} {\bibfnamefont {V.}~\bibnamefont {Pardo}}, \ and\
  \bibinfo {author} {\bibfnamefont {W.~E.}\ \bibnamefont {Pickett}},\ }\href
  {\doibase 10.1103/PhysRevLett.103.016402} {\bibfield  {journal} {\bibinfo
  {journal} {Phys. Rev. Lett.}\ }\textbf {\bibinfo {volume} {103}},\ \bibinfo
  {pages} {016402} (\bibinfo {year} {2009})}\BibitemShut {NoStop}%
\bibitem [{\citenamefont {Montambaux}\ \emph
  {et~al.}(2009{\natexlab{a}})\citenamefont {Montambaux}, \citenamefont
  {Pi\'echon}, \citenamefont {Fuchs},\ and\ \citenamefont
  {Goerbig}}]{PhysRevB.80.153412}%
  \BibitemOpen
  \bibfield  {author} {\bibinfo {author} {\bibfnamefont {G.}~\bibnamefont
  {Montambaux}}, \bibinfo {author} {\bibfnamefont {F.}~\bibnamefont
  {Pi\'echon}}, \bibinfo {author} {\bibfnamefont {J.-N.}\ \bibnamefont
  {Fuchs}}, \ and\ \bibinfo {author} {\bibfnamefont {M.~O.}\ \bibnamefont
  {Goerbig}},\ }\href {\doibase 10.1103/PhysRevB.80.153412} {\bibfield
  {journal} {\bibinfo  {journal} {Phys. Rev. B}\ }\textbf {\bibinfo {volume}
  {80}},\ \bibinfo {pages} {153412} (\bibinfo {year}
  {2009}{\natexlab{a}})}\BibitemShut {NoStop}%
\bibitem [{\citenamefont {Montambaux}\ \emph
  {et~al.}(2009{\natexlab{b}})\citenamefont {Montambaux}, \citenamefont
  {Pi{\'e}chon}, \citenamefont {Fuchs},\ and\ \citenamefont
  {Goerbig}}]{universal}%
  \BibitemOpen
  \bibfield  {author} {\bibinfo {author} {\bibfnamefont {G.}~\bibnamefont
  {Montambaux}}, \bibinfo {author} {\bibfnamefont {F.}~\bibnamefont
  {Pi{\'e}chon}}, \bibinfo {author} {\bibfnamefont {J.-N.}\ \bibnamefont
  {Fuchs}}, \ and\ \bibinfo {author} {\bibfnamefont {M.~O.}\ \bibnamefont
  {Goerbig}},\ }\href@noop {} {\bibfield  {journal} {\bibinfo  {journal} {Eur.
  Phys. J. B}\ }\textbf {\bibinfo {volume} {72}},\ \bibinfo {pages} {509}
  (\bibinfo {year} {2009}{\natexlab{b}})}\BibitemShut {NoStop}%
\bibitem [{\citenamefont {Banerjee}\ and\ \citenamefont
  {Pickett}(2012)}]{PhysRevB.86.075124}%
  \BibitemOpen
  \bibfield  {author} {\bibinfo {author} {\bibfnamefont {S.}~\bibnamefont
  {Banerjee}}\ and\ \bibinfo {author} {\bibfnamefont {W.~E.}\ \bibnamefont
  {Pickett}},\ }\href {\doibase 10.1103/PhysRevB.86.075124} {\bibfield
  {journal} {\bibinfo  {journal} {Phys. Rev. B}\ }\textbf {\bibinfo {volume}
  {86}},\ \bibinfo {pages} {075124} (\bibinfo {year} {2012})}\BibitemShut
  {NoStop}%
\bibitem [{\citenamefont {Hohenberg}\ and\ \citenamefont {Kohn}(1964)}]{DFT}%
  \BibitemOpen
  \bibfield  {author} {\bibinfo {author} {\bibfnamefont {P.}~\bibnamefont
  {Hohenberg}}\ and\ \bibinfo {author} {\bibfnamefont {W.}~\bibnamefont
  {Kohn}},\ }\href {\doibase 10.1103/PhysRev.136.B864} {\bibfield  {journal}
  {\bibinfo  {journal} {Phys. Rev.}\ }\textbf {\bibinfo {volume} {136}},\
  \bibinfo {pages} {B864} (\bibinfo {year} {1964})}\BibitemShut {NoStop}%
\bibitem [{\citenamefont {Giannozzi}\ and\ \citenamefont {\textit{et
  al.}}(2009)}]{QE}%
  \BibitemOpen
  \bibfield  {author} {\bibinfo {author} {\bibfnamefont {P.}~\bibnamefont
  {Giannozzi}}\ and\ \bibinfo {author} {\bibnamefont {\textit{et al.}}},\
  }\href {http://stacks.iop.org/0953-8984/21/i=39/a=395502} {\bibfield
  {journal} {\bibinfo  {journal} {J. Phys.: Condens. Matter}\ }\textbf
  {\bibinfo {volume} {21}},\ \bibinfo {pages} {395502} (\bibinfo {year}
  {2009})},\ \bibinfo {note}
  {\url{http://www.quantum-espresso.org}}\BibitemShut {NoStop}%
\bibitem [{\citenamefont {Kresse}\ and\ \citenamefont
  {Furthm\"{u}ller}(1996)}]{VASP}%
  \BibitemOpen
  \bibfield  {author} {\bibinfo {author} {\bibfnamefont {G.}~\bibnamefont
  {Kresse}}\ and\ \bibinfo {author} {\bibfnamefont {J.}~\bibnamefont
  {Furthm\"{u}ller}},\ }\href@noop {} {\bibfield  {journal} {\bibinfo
  {journal} {Comput. Mater. Sci.}\ }\textbf {\bibinfo {volume} {6}},\ \bibinfo
  {pages} {15} (\bibinfo {year} {1996})},\ \bibinfo {note}
  {\url{http://www.vasp.at}}\BibitemShut {NoStop}%
\bibitem [{\citenamefont {Perdew}\ \emph {et~al.}(1996)\citenamefont {Perdew},
  \citenamefont {Burke},\ and\ \citenamefont {Ernzerhof}}]{PBE}%
  \BibitemOpen
  \bibfield  {author} {\bibinfo {author} {\bibfnamefont {J.~P.}\ \bibnamefont
  {Perdew}}, \bibinfo {author} {\bibfnamefont {K.}~\bibnamefont {Burke}}, \
  and\ \bibinfo {author} {\bibfnamefont {M.}~\bibnamefont {Ernzerhof}},\
  }\href@noop {} {\bibfield  {journal} {\bibinfo  {journal} {Phys.\ Rev.
  Lett.}\ }\textbf {\bibinfo {volume} {77}},\ \bibinfo {pages} {3865} (\bibinfo
  {year} {1996})}\BibitemShut {NoStop}%
\bibitem [{\citenamefont {Anisimov}\ \emph {et~al.}(1991)\citenamefont
  {Anisimov}, \citenamefont {Zaanen},\ and\ \citenamefont
  {Andersen}}]{HubbardU}%
  \BibitemOpen
  \bibfield  {author} {\bibinfo {author} {\bibfnamefont {V.~I.}\ \bibnamefont
  {Anisimov}}, \bibinfo {author} {\bibfnamefont {J.}~\bibnamefont {Zaanen}}, \
  and\ \bibinfo {author} {\bibfnamefont {O.~K.}\ \bibnamefont {Andersen}},\
  }\href {\doibase 10.1103/PhysRevB.44.943} {\bibfield  {journal} {\bibinfo
  {journal} {Phys. Rev. B}\ }\textbf {\bibinfo {volume} {44}},\ \bibinfo
  {pages} {943} (\bibinfo {year} {1991})}\BibitemShut {NoStop}%
\bibitem [{\citenamefont {Dudarev}\ \emph {et~al.}(1998)\citenamefont
  {Dudarev}, \citenamefont {Botton}, \citenamefont {Savrasov}, \citenamefont
  {Humphreys},\ and\ \citenamefont {Sutton}}]{HubbardU2}%
  \BibitemOpen
  \bibfield  {author} {\bibinfo {author} {\bibfnamefont {S.~L.}\ \bibnamefont
  {Dudarev}}, \bibinfo {author} {\bibfnamefont {G.~A.}\ \bibnamefont {Botton}},
  \bibinfo {author} {\bibfnamefont {S.~Y.}\ \bibnamefont {Savrasov}}, \bibinfo
  {author} {\bibfnamefont {C.~J.}\ \bibnamefont {Humphreys}}, \ and\ \bibinfo
  {author} {\bibfnamefont {A.~P.}\ \bibnamefont {Sutton}},\ }\href {\doibase
  10.1103/PhysRevB.57.1505} {\bibfield  {journal} {\bibinfo  {journal} {Phys.
  Rev. B}\ }\textbf {\bibinfo {volume} {57}},\ \bibinfo {pages} {1505}
  (\bibinfo {year} {1998})}\BibitemShut {NoStop}%
\bibitem [{\citenamefont {Marzari}\ and\ \citenamefont
  {Vanderbilt}(1997)}]{wannier1}%
  \BibitemOpen
  \bibfield  {author} {\bibinfo {author} {\bibfnamefont {N.}~\bibnamefont
  {Marzari}}\ and\ \bibinfo {author} {\bibfnamefont {D.}~\bibnamefont
  {Vanderbilt}},\ }\href {\doibase 10.1103/PhysRevB.56.12847} {\bibfield
  {journal} {\bibinfo  {journal} {Phys. Rev. B}\ }\textbf {\bibinfo {volume}
  {56}},\ \bibinfo {pages} {12847} (\bibinfo {year} {1997})}\BibitemShut
  {NoStop}%
\bibitem [{\citenamefont {Souza}\ \emph {et~al.}(2001)\citenamefont {Souza},
  \citenamefont {Marzari},\ and\ \citenamefont {Vanderbilt}}]{wannier2}%
  \BibitemOpen
  \bibfield  {author} {\bibinfo {author} {\bibfnamefont {I.}~\bibnamefont
  {Souza}}, \bibinfo {author} {\bibfnamefont {N.}~\bibnamefont {Marzari}}, \
  and\ \bibinfo {author} {\bibfnamefont {D.}~\bibnamefont {Vanderbilt}},\
  }\href {\doibase 10.1103/PhysRevB.65.035109} {\bibfield  {journal} {\bibinfo
  {journal} {Phys. Rev. B}\ }\textbf {\bibinfo {volume} {65}},\ \bibinfo
  {pages} {035109} (\bibinfo {year} {2001})}\BibitemShut {NoStop}%
\bibitem [{\citenamefont {Mostofi}\ \emph {et~al.}(2008)\citenamefont
  {Mostofi}, \citenamefont {Yates}, \citenamefont {Lee}, \citenamefont {Souza},
  \citenamefont {Vanderbilt},\ and\ \citenamefont {Marzari}}]{wannier90}%
  \BibitemOpen
  \bibfield  {author} {\bibinfo {author} {\bibfnamefont {A.~A.}\ \bibnamefont
  {Mostofi}}, \bibinfo {author} {\bibfnamefont {J.~R.}\ \bibnamefont {Yates}},
  \bibinfo {author} {\bibfnamefont {Y.-S.}\ \bibnamefont {Lee}}, \bibinfo
  {author} {\bibfnamefont {I.}~\bibnamefont {Souza}}, \bibinfo {author}
  {\bibfnamefont {D.}~\bibnamefont {Vanderbilt}}, \ and\ \bibinfo {author}
  {\bibfnamefont {N.}~\bibnamefont {Marzari}},\ }\href@noop {} {\bibfield
  {journal} {\bibinfo  {journal} {Comput. Phys. Commun.}\ }\textbf {\bibinfo
  {volume} {178}},\ \bibinfo {pages} {685} (\bibinfo {year}
  {2008})}\BibitemShut {NoStop}%
\bibitem [{\citenamefont {Goodenough}(1971)}]{goodenough}%
  \BibitemOpen
  \bibfield  {author} {\bibinfo {author} {\bibfnamefont {J.~B.}\ \bibnamefont
  {Goodenough}},\ }\href@noop {} {\bibfield  {journal} {\bibinfo  {journal} {J.
  Solid State Chem.}\ }\textbf {\bibinfo {volume} {3}},\ \bibinfo {pages} {490}
  (\bibinfo {year} {1971})}\BibitemShut {NoStop}%
\bibitem [{\citenamefont {Yates}\ \emph {et~al.}(2007)\citenamefont {Yates},
  \citenamefont {Wang}, \citenamefont {Vanderbilt},\ and\ \citenamefont
  {Souza}}]{Yates}%
  \BibitemOpen
  \bibfield  {author} {\bibinfo {author} {\bibfnamefont {J.~R.}\ \bibnamefont
  {Yates}}, \bibinfo {author} {\bibfnamefont {X.}~\bibnamefont {Wang}},
  \bibinfo {author} {\bibfnamefont {D.}~\bibnamefont {Vanderbilt}}, \ and\
  \bibinfo {author} {\bibfnamefont {I.}~\bibnamefont {Souza}},\ }\href
  {\doibase 10.1103/PhysRevB.75.195121} {\bibfield  {journal} {\bibinfo
  {journal} {Phys. Rev. B}\ }\textbf {\bibinfo {volume} {75}},\ \bibinfo
  {pages} {195121} (\bibinfo {year} {2007})}\BibitemShut {NoStop}%
\bibitem [{\citenamefont {Wang}\ \emph {et~al.}(2006)\citenamefont {Wang},
  \citenamefont {Yates}, \citenamefont {Souza},\ and\ \citenamefont
  {Vanderbilt}}]{WangXinjie}%
  \BibitemOpen
  \bibfield  {author} {\bibinfo {author} {\bibfnamefont {X.}~\bibnamefont
  {Wang}}, \bibinfo {author} {\bibfnamefont {J.~R.}\ \bibnamefont {Yates}},
  \bibinfo {author} {\bibfnamefont {I.}~\bibnamefont {Souza}}, \ and\ \bibinfo
  {author} {\bibfnamefont {D.}~\bibnamefont {Vanderbilt}},\ }\href {\doibase
  10.1103/PhysRevB.74.195118} {\bibfield  {journal} {\bibinfo  {journal} {Phys.
  Rev. B}\ }\textbf {\bibinfo {volume} {74}},\ \bibinfo {pages} {195118}
  (\bibinfo {year} {2006})}\BibitemShut {NoStop}%
\bibitem [{\citenamefont {Nagaosa}\ \emph {et~al.}(2010)\citenamefont
  {Nagaosa}, \citenamefont {Sinova}, \citenamefont {Onoda}, \citenamefont
  {MacDonald},\ and\ \citenamefont {Ong}}]{ahc}%
  \BibitemOpen
  \bibfield  {author} {\bibinfo {author} {\bibfnamefont {N.}~\bibnamefont
  {Nagaosa}}, \bibinfo {author} {\bibfnamefont {J.}~\bibnamefont {Sinova}},
  \bibinfo {author} {\bibfnamefont {S.}~\bibnamefont {Onoda}}, \bibinfo
  {author} {\bibfnamefont {A.~H.}\ \bibnamefont {MacDonald}}, \ and\ \bibinfo
  {author} {\bibfnamefont {N.~P.}\ \bibnamefont {Ong}},\ }\href {\doibase
  10.1103/RevModPhys.82.1539} {\bibfield  {journal} {\bibinfo  {journal} {Rev.
  Mod. Phys.}\ }\textbf {\bibinfo {volume} {82}},\ \bibinfo {pages} {1539}
  (\bibinfo {year} {2010})}\BibitemShut {NoStop}%
\bibitem [{\citenamefont {L{\'o}pez~Sancho}\ \emph {et~al.}(1984)\citenamefont
  {L{\'o}pez~Sancho}, \citenamefont {L{\'o}pez~Sancho},\ and\ \citenamefont
  {Rubio}}]{lopez}%
  \BibitemOpen
  \bibfield  {author} {\bibinfo {author} {\bibfnamefont {M.~P.}\ \bibnamefont
  {L{\'o}pez~Sancho}}, \bibinfo {author} {\bibfnamefont {J.~M.}\ \bibnamefont
  {L{\'o}pez~Sancho}}, \ and\ \bibinfo {author} {\bibfnamefont
  {J.}~\bibnamefont {Rubio}},\ }\href@noop {} {\bibfield  {journal} {\bibinfo
  {journal} {J. Phys. F}\ }\textbf {\bibinfo {volume} {14}},\ \bibinfo {pages}
  {1205} (\bibinfo {year} {1984})}\BibitemShut {NoStop}%
\bibitem [{\citenamefont {L{\'o}pez~Sancho}\ \emph {et~al.}(1985)\citenamefont
  {L{\'o}pez~Sancho}, \citenamefont {L{\'o}pez~Sancho},\ and\ \citenamefont
  {Rubio}}]{lopez2}%
  \BibitemOpen
  \bibfield  {author} {\bibinfo {author} {\bibfnamefont {M.~P.}\ \bibnamefont
  {L{\'o}pez~Sancho}}, \bibinfo {author} {\bibfnamefont {J.~M.}\ \bibnamefont
  {L{\'o}pez~Sancho}}, \ and\ \bibinfo {author} {\bibfnamefont
  {J.}~\bibnamefont {Rubio}},\ }\href
  {http://stacks.iop.org/0305-4608/15/i=4/a=009} {\bibfield  {journal}
  {\bibinfo  {journal} {J. Phys. F}\ }\textbf {\bibinfo {volume} {15}},\
  \bibinfo {pages} {851} (\bibinfo {year} {1985})}\BibitemShut {NoStop}%
\bibitem [{\citenamefont {Solovyev}\ \emph {et~al.}(2007)\citenamefont
  {Solovyev}, \citenamefont {Pchelkina},\ and\ \citenamefont
  {Anisimov}}]{downfold1}%
  \BibitemOpen
  \bibfield  {author} {\bibinfo {author} {\bibfnamefont {I.~V.}\ \bibnamefont
  {Solovyev}}, \bibinfo {author} {\bibfnamefont {Z.~V.}\ \bibnamefont
  {Pchelkina}}, \ and\ \bibinfo {author} {\bibfnamefont {V.~I.}\ \bibnamefont
  {Anisimov}},\ }\href {\doibase 10.1103/PhysRevB.75.045110} {\bibfield
  {journal} {\bibinfo  {journal} {Phys. Rev. B}\ }\textbf {\bibinfo {volume}
  {75}},\ \bibinfo {pages} {045110} (\bibinfo {year} {2007})}\BibitemShut
  {NoStop}%
\bibitem [{\citenamefont {Huang}\ \emph {et~al.}(2013)\citenamefont {Huang},
  \citenamefont {Duan},\ and\ \citenamefont {Liu}}]{downfold2}%
  \BibitemOpen
  \bibfield  {author} {\bibinfo {author} {\bibfnamefont {H.}~\bibnamefont
  {Huang}}, \bibinfo {author} {\bibfnamefont {W.}~\bibnamefont {Duan}}, \ and\
  \bibinfo {author} {\bibfnamefont {Z.}~\bibnamefont {Liu}},\ }\href
  {http://stacks.iop.org/1367-2630/15/i=2/a=023004} {\bibfield  {journal}
  {\bibinfo  {journal} {New J. Phys.}\ }\textbf {\bibinfo {volume} {15}},\
  \bibinfo {pages} {023004} (\bibinfo {year} {2013})}\BibitemShut {NoStop}%
\bibitem [{Note1()}]{Note1}%
  \BibitemOpen
  \bibinfo {note} {\label {fn}See Supplementary Material
  http://link.aps.org/supplemental/xxx for more computation details, the
  derivation of the $\protect \mathbf {k}\cdot \protect \mathbf {p}$ model, and
  a detailed analysis of the electronic structure.}\BibitemShut {Stop}%
\bibitem [{\citenamefont {Cai}\ \emph {et~al.}(2013)\citenamefont {Cai},
  \citenamefont {Li}, \citenamefont {Wang}, \citenamefont {Sheng},
  \citenamefont {Feng},\ and\ \citenamefont {Gong}}]{Cai}%
  \BibitemOpen
  \bibfield  {author} {\bibinfo {author} {\bibfnamefont {T.-Y.}\ \bibnamefont
  {Cai}}, \bibinfo {author} {\bibfnamefont {X.}~\bibnamefont {Li}}, \bibinfo
  {author} {\bibfnamefont {F.}~\bibnamefont {Wang}}, \bibinfo {author}
  {\bibfnamefont {J.}~\bibnamefont {Sheng}}, \bibinfo {author} {\bibfnamefont
  {J.}~\bibnamefont {Feng}}, \ and\ \bibinfo {author} {\bibfnamefont {C.-D.}\
  \bibnamefont {Gong}},\ }\href@noop {} {\bibfield  {journal} {\bibinfo
  {journal} {{arXiv: 1310.2471}}\ } (\bibinfo {year} {2013})}\BibitemShut
  {NoStop}%
\end{thebibliography}
\providecommand{\noopsort}[1]{}\providecommand{\singleletter}[1]{#1}%

\end{document}